\documentclass[twocolumn,twocolappendix]{aastex631}

\usepackage{cancel}
\usepackage[dvipsnames]{xcolor}

\begin{document}

\title{Integrated Mass Loss for Very Metal-poor Stars: \\Asterosesimic Red Giant Masses of K2 Globular Cluster NGC 5897}


\author[0000-0002-1663-0707]{Csilla Kalup}
\affiliation{Konkoly Observatory, HUN-REN CSFK, Konkoly-Thege Mikl\'os \'ut 15-17, H-1121, Budapest, Hungary}
\affiliation{CSFK, MTA Centre of Excellence, Konkoly-Thege Mikl\'os \'ut 15-17, H-1121, Budapest, Hungary}
\affiliation{E\"otv\"os Lor\'and University, Institute of Physics and Astronomy, P\'azm\'any P\'eter s\'et\'any 1/A, H-1117, Budapest, Hungary}

\author[0000-0002-8159-1599]{László Molnár}
\affiliation{Konkoly Observatory, HUN-REN CSFK, Konkoly-Thege Mikl\'os \'ut 15-17, H-1121, Budapest, Hungary}
\affiliation{CSFK, MTA Centre of Excellence, Konkoly-Thege Mikl\'os \'ut 15-17, H-1121, Budapest, Hungary}
\affiliation{E\"otv\"os Lor\'and University, Institute of Physics and Astronomy, P\'azm\'any P\'eter s\'et\'any 1/A, H-1117, Budapest, Hungary}

\author[0000-0003-0929-6541]{Madeline Howell}
\affiliation{Department of Astronomy, The Ohio State University, 140 W. 18th Ave., Columbus, OH 43210, USA}
\affiliation{Center for Cosmology and Astroparticle Physics (CCAPP), The Ohio State University, 191 West Woodruff Avenue, Columbus, OH 43210, USA}

\author[0000-0002-8585-4544]{Attila Bódi}
\affiliation{Department of Astrophysical Sciences, Princeton University, Peyton Hall, 4 Ivy Lane, Princeton, NJ 08544, USA}

\author[0000-0001-5449-2467]{András Pál}
\affiliation{Konkoly Observatory, HUN-REN CSFK, Konkoly-Thege Mikl\'os \'ut 15-17, H-1121, Budapest, Hungary}
\affiliation{CSFK, MTA Centre of Excellence, Konkoly-Thege Mikl\'os \'ut 15-17, H-1121, Budapest, Hungary}
\affiliation{E\"otv\"os Lor\'and University, Institute of Physics and Astronomy, P\'azm\'any P\'eter s\'et\'any 1/A, H-1117, Budapest, Hungary}






\begin{abstract}

Mass loss in low-mass stars during the red giant branch (RGB) and early asymptotic giant branch (EAGB) phases plays a key role in shaping stellar evolution, yet its dependence on stellar parameters such as metallicity remains poorly constrained, with observational studies yielding conflicting trends.
We present the first asteroseismic analysis of RGB and EAGB stars in NGC 5897, the most distant and metal-poor globular cluster observed by the \textit{Kepler} space telescope during the K2 mission. We detected solar-like oscillations and derived the frequency of maximum power excess, $\nu_{\rm max}$, for 20 RGB and 6 EAGB stars.
Using asteroseismic scaling relations, we derived mean masses of $\overline{M}_{\rm RGB}=0.74\pm0.01\,M_{\odot}$ and $\overline{M}_{\rm EAGB}=0.65\pm 0.03\,M_{\odot}$. The inferred integrated mass loss between the two phases is $\Delta M_{\rm RGB-EAGB}=0.08\pm 0.03\,M_{\odot}$.
We present an updated mass-loss--metallicity relation for Type~I globular clusters, extending it to the very metal-poor regime and supporting decreasing integrated RGB mass loss with decreasing metallicity.

\end{abstract}




\section{Introduction} \label{sec:intro}



Globular clusters are among the oldest and most metal-poor imprints of the early formation of the Milky Way. Besides their role in Galactic archeology in tracing the formation history of the Galaxy \citep{Forbes-2018,Kruijssen-2019,Massari-2019}, they also serve as cosmic laboratories to study both stellar evolution and the pulsation of low-mass stars at various evolutionary phases \citep[see, e.g,][for some examples]{Catelan-2009,Gratton-2010,Kalirai-2010,Jurcsik-2015,Bastian-2018}. The color--magnitude diagram (CMD) of these clusters allows us to photometrically distinguish different evolutionary stages, with particular regard to stars on the red giant branch (RGB), the horizontal branch (HB), the asymptotic giant branch (AGB), especially its early-AGB (EAGB) phase, before the thermally pulsing stage.

One of the most puzzling questions in stellar evolution is how stars lose mass during their evolution. 
Although massive stars lose the most significant amount of mass \citep{Antoniadis2024}, mass loss via stellar winds play an important role in low-mass stars too \citep{Willson2000}. 
Whilst mass loss primarily dominates the AGB phase \citep{Hofner2018}, more mass can be lost during the RGB phase in stars with $M\lesssim1\,M_{\odot}$ due to their longer lifespan \citep{McDonald2015}. 
Even though mass loss is such an important factor, it is still poorly understood and lacks of any comprehensive theoretical description. In stellar modeling, only simple prescriptions such as the \citet{Reimers-1975} scheme and its semi-empirical variants \citep[e.g.][]{Schroder-2005} are applied, although observations suggest more complex scenarios, including metallicity dependency. 
Recent measurements have revealed a clear contradiction of mass-loss--metallicity trends: while open clusters and stars with $-0.8<\rm[Fe/H]<+0.4$ show increasing mass loss with decreasing metallicity \citep{Miglio-2012,Handberg-2017,Brogaard-2024,Reyes-2025,Yaguang-2025}, metal-poor samples indicate a decrease with decreasing metallicity \citep{Tailo2021,Maddy2025}. This conflict highlights the need for additional observational constraints on RGB mass loss and further theoretical development.

Asteroseismology of red giants in globular clusters is one avenue to measure mass loss, particularly for old and metal-poor stars. 
RGB and EAGB stars exhibit solar-like oscillations, and these stochastically driven pressure-mode oscillations can probe the stellar interior \citep{Aerts-2021}. Through asteroseismology, we can characterize global seismic parameters of an oscillating star's power spectrum: $\nu_{\mathrm{max}}$, the frequency of the maximum acoustic power, is related to surface gravity and effective temperature \citep{Kjeldsen-Bedding-1995}, while $\Delta\nu$, the large frequency separation correlates with mean density \citep{Tassoul1980}. Combining these quantities with physical properties and/or stellar evolutionary models, it is possible to infer seismic masses, ages or radii for these stars \citep{Huber2011,Pinsonneault2018}. 
However, detailed analysis of stellar pulsations requires high-precision, uninterrupted, long-term measurements. During the K2 mission of the \textit{Kepler} space telescope \citep{Howell-2014}, eight globular clusters were observed, and recent studies already demonstrated their asteroseismic potential. This is especially true for M4 \citep{Miglio2016,Wallace-2019,Wallace-2020,Tailo2022,Maddy2022,Molnar-2024}, but M80, M9 and M19 have also been investigated \citep{Molnar2023,Maddy2024,Maddy2025}.

In this paper, we present the first asteroseismic analysis of NGC 5897, which is the fifth K2 cluster to be investigated, and the most distant and metal-poor among them. We examine the RGB and EAGB populations of NGC 5897 in order to extend and constrain the current K2 globular cluster mass-loss--metallicity relation \citep{Maddy2025} at the very metal-poor end. In Section \ref{sect:data}, we introduce our RGB and EAGB samples. 
In Section \ref{sect:reduction}, we describe our photometric pipeline and the post-processing steps we applied. In Section \ref{sect:scaling_rel}, we derive the seismic masses of our samples. In Section \ref{sect:masses}, we present the average RGB and EAGB masses. In Section \ref{sect:conclusions}, we discuss the results in light of the metallicity dependence of the integrated mass loss.

\section{K2 red giant sample of NGC 5897}
\label{sect:data}

NGC 5897 is an inner halo globular cluster associated with the Gaia--Enceladus merging event \citep{Massari-2019}. It has a metallicity of [Fe/H] $= -2.04  \pm 0.01$ (stat) $\pm 0.15$ (sys), based on spectroscopic measurements \citep{Koch2014}. The cluster is well-known for unusually long period RR Lyrae stars, which also supports the low-metallicity value of the cluster \citep{Clement-2001}.

To identify the RGB and EAGB samples for NGC 5897, we constructed its CMD using Johnson $B$ and $V$ photometry collected by \cite{Stetson2019}, after cross-matching each star with their membership probabilities published by \cite{Vasiliev2021} 
with a limit of $\geq99\%$. We identified a suitable brightness range ($15.8<V<14.3$\,mag) for our initial red giant sample (hereafter, RG sample) to search for solar-like oscillations. 
While brighter stars transition into semi-regular variables, where the K2 data becomes too short to resolve the oscillations, fainter stars fall below the sensitivity limit of the K2 data to detect low-amplitude signals, as mode amplitudes decrease with decreasing luminosity \citep{Stello2011}. The selected region is in a similar absolute brightness range as the seismically analyzed upper RGB sections of other K2 clusters.

\begin{figure}
\includegraphics[width=1.0\columnwidth]{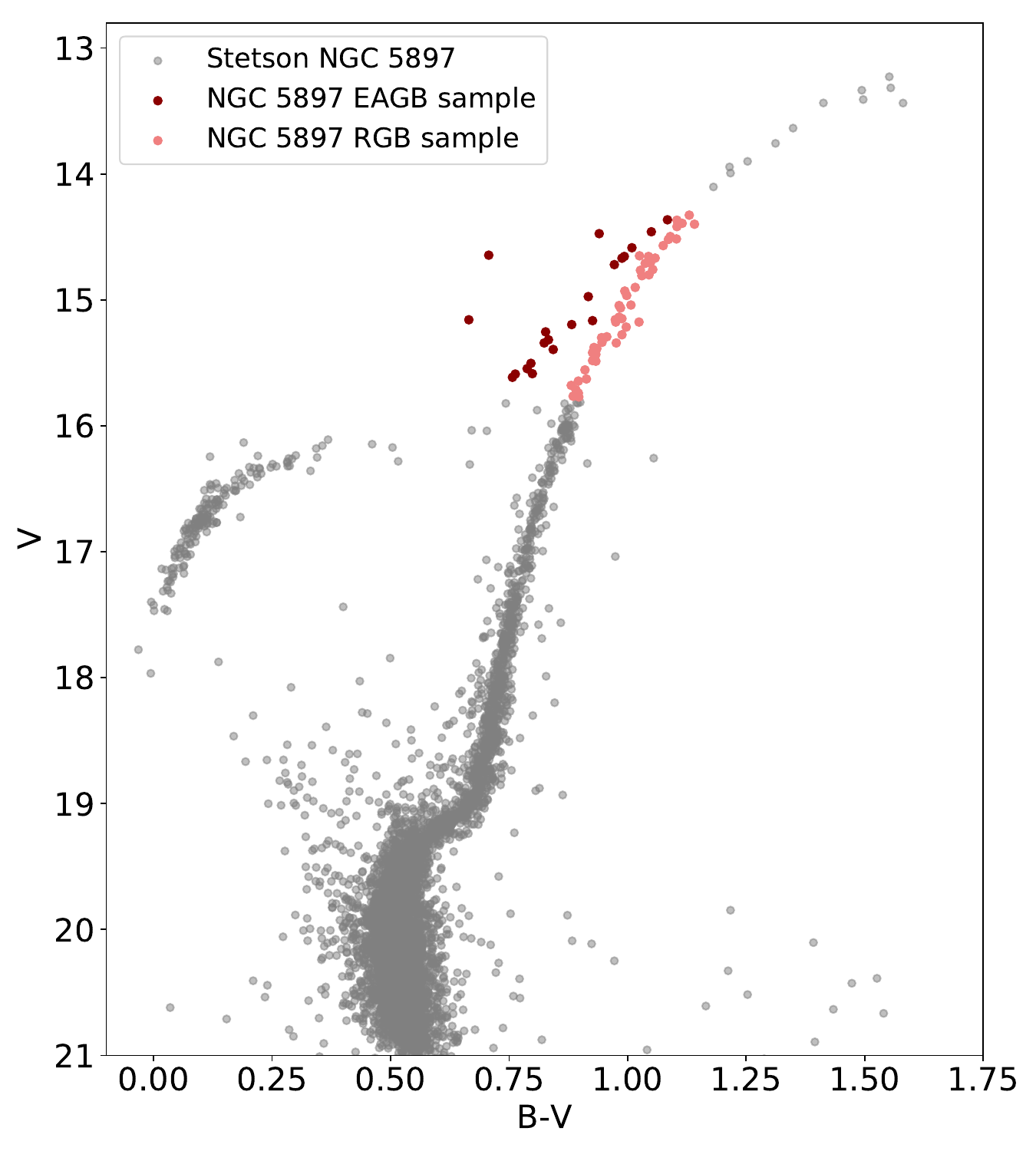}
\caption{Color-magnitude diagram of NGC 5897 based on Johnson $B$ and $V$ photometry by \cite{Stetson2019}. The colored points represent the initial red giant sample we searched for solar-like oscillations. The 21 dark red dots correspond to early-AGB stars, while the 49 light pink points show RGB stars.
\label{fig:BV_CMD}}
\end{figure}

Figure \ref{fig:BV_CMD} shows how we classified our RG sample into RGB (dark red) and EAGB stars (pink) by visual inspection based on their position on the CMD. 
Each star in our RG sample was designated 
by their evolutionary stage, following a number corresponding to their $V$ brightness, where a lower number refers to a brighter star (e.g. RGB10, see more in Table~\ref{tab:results}). In order to confirm these evolutionary stages, we also created CMDs using the Stetson $V$ and $I$ magnitudes \citep{Stetson2019}, \textit{Gaia} $G$, $G_{\rm BP}$ and $G_{\rm RP}$, as well as 2MASS $K$ photometry \citep{Skrutskie-2006}. More details on this can be found in Appendix~\ref{sect:AppendixA}. Overall, we find 21 EAGB and 49 RGB stars in the initial RG sample, from which five stars did not fall on active part of the CCD. We also excluded 11 more stars after searching for contamination from nearby stars.

\section{Data reduction}
\label{sect:reduction}

The NGC 5897 cluster was observed during Campaign 15 of the K2 mission of the \textit{Kepler} space telescope, from August 23 to November 20, 2017, with a total baseline of 88 days.
The images were collected in long-cadence mode with 29.41 min sampling, using a 90$\times$90 pixel square box, corresponding to a 6'$\times$6' field-of-view. The data we use in this paper were proposed through K2 GO programs 15009, 15020, 15021, 15058, 15083, 15092, 15903 and spans the EPIC (K2 Ecliptic Plane Input Catalog) IDs 200194922 to 200194957.

\subsection{Differential photometry}
\label{sect:diffphot}

We applied aperture photometry to differential images in order to handle overlapping point spread functions (PSFs) of crowded stellar fields. 
For this we used the FITSH software package \citep{PAl2012} and our external \texttt{bash} scripts, which have been successfully utilized in many other K2 time-series analyses for moving targets of the Solar System \citep{FarkasTakacs2017, Molnar2018a, Marton2020, Szabo2020, Kalup2021} or stellar variable sources \citep{Plachy2017, Vida2017, Molnar2023,Molnar-2024}. 

Briefly, we queried the Target Pixel Files of the cluster plus some bright, nearby stars from the Mikulski Archive for Space Telescopes\footnote{\url{https://archive.stsci.edu/}} (MAST) database to assemble a mosaic image \citep{K2-C15}. Then, we calculated the astrometric solutions for these frames, identifying the brightest stars using the USNO (United States Naval Observatory) catalog\footnote{\url{https://irsa.ipac.caltech.edu/Missions/usno.html}}, based on the initial solution for the full field images of the campaign, to register them into the same reference system. We created a master image from a few dozen images and subtracted it from all the frames to only retain signals belonging to moving and variable features. To reduce the effects of sharp PSFs due to the registration, interpolation, and master image subtraction, we also enlarged each frame by $\sim$3 times, distributing the original flux level among more pixels. This can cause smoother photometry, and allows finer control over the apertures. We note that in the denser regions of the field we used tight apertures, which produced cleaner light curves but did not fully cover the PSF edges, leading to underestimated amplitudes. Amplitudes were not used in this paper, nor do we recommend it.

\subsection{Postprocessing}
\label{sect:postproc}

The raw light curves 
are heavily affected by various systematics and instrumental signals. We combined the K2SC package \citep{Aigrain2016} with a phase dispersion minimization optimized trend fitter \citep{Bodi2022} in order to 
remove instrumental systematics and slow trends, while keeping the quasi-periodic astrophysical signal intact. For more details 
we recommend the K2 data reduction section of M80 by \cite{Molnar2023}. For an independent check, we also tested the Self-Flat Field Correction method within the \texttt{lightkurve} package, introduced by \citet{Vanderburg-2014}, following suggestions by \cite{Maddy2022, Maddy2024, Maddy2025}, and we found general agreement between the calculated $\nu_{\rm max}$ values. The corrected light curves are available online, as detailed in Appendix \ref{sect:AppendixB}.

\subsection{Power excess detections}

\begin{figure}
\includegraphics[width=1.0\columnwidth]{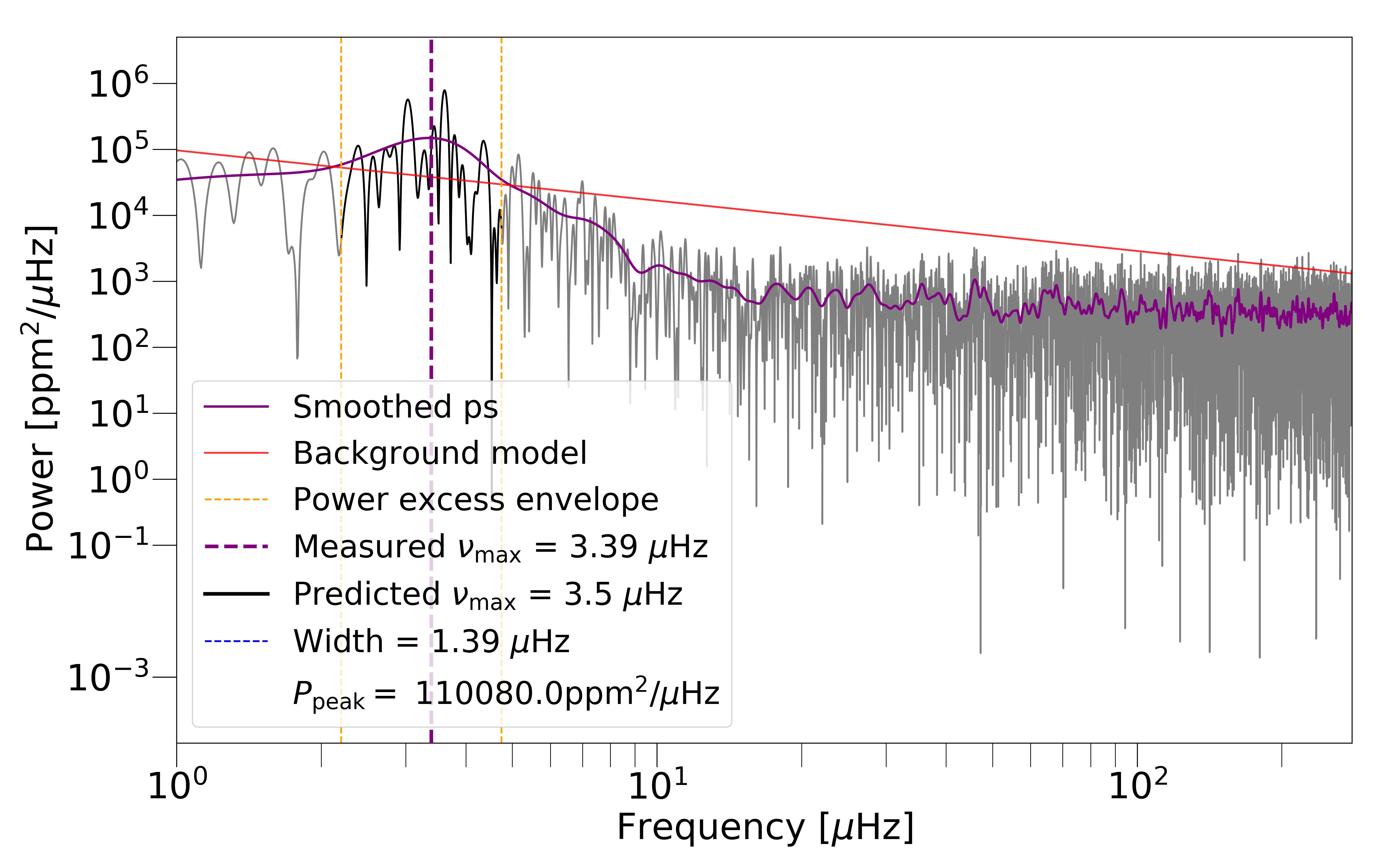}
\caption{Power spectrum of RGB17. The gray curve represents the power density spectrum, purple is the smoothed spectrum. The power excess of the solar-like oscillations is between the two yellow lines, while the red line is the linear background fit within this range. The $\nu_{\rm max}$ estimated by the pyMON code is indicated as a vertical dashed purple line.
\label{fig:spec}}
\end{figure}

We applied multiple circular apertures with increasing radii to each star during photometry. After the post-processing steps, we visually inspected each light curve by searching for signs of the power excess in the Lomb-Scargle periodogram \citep{VanderPlas2018}. Then, we chose the best aperture that could maximize the signal in the power spectrum. We inferred $\nu_{\rm max}$ values using the \texttt{pyMON}\footnote{The pyMON repository can be accessed here: \url{https://github.com/maddyhowell/pyMON}} code, which was developed specially for low-frequency solar-like oscillators in K2 clusters -- see a detailed description in \cite{Maddy2025} through the cases of K2 globular clusters M9 and M19, and an example spectrum in Figure \ref{fig:spec}. Overall, we were able to infer $\nu_{\rm max}$ for 6 EAGB and 20 RGB stars. The derived seismic parameters are listed in Table \ref{tab:results}.

\begin{deluxetable*}{lccccccccc}
\tablecaption{Derived parameters for RGB and EAGB stars in NGC 5897 with successful power excess detections. We provide inferred and calculated stellar parameters such as the frequency of maximum power excess ($\nu_{\rm max}$), photometric temperature ($T_{\rm eff}$), luminosity ($L$), stellar mass ($M$) and their corresponding uncertainties.
\label{tab:results}}
\tablehead{
\colhead{ID} & \colhead{Gaia DR3 source ID} & \colhead{$\nu_{max}$ [$\mu$Hz]}  & \colhead{$\sigma_{\nu_{max}}$ [$\mu$Hz]}& \colhead{T$_{\rm eff}$ [K]} & \colhead{$\sigma_{T_{\rm eff}}$ [K]} & \colhead{L [L$_{\odot}$]} & \colhead{$\sigma_L$ [L$_{\odot}$]}  & \colhead{M [M$_{\odot}$]} & \colhead{$\sigma_M$ [M$_{\odot}$]}}
\startdata
EAGB04	&	6252665788119836672	&	2.709	&	0.044	&	4660	&	61	& 334	&	18	&	0.620	&	0.045	\\ 
EAGB10	&	6252664108792246400	&	7.231	&	0.301	&	5058	&	69	& 175	&	9	&	0.649	&	0.053	\\ 
EAGB11	&	6252666548333610624	&	6.593	&	0.082	&	4797    &	70	& 186	&	10	&	0.761	&	0.058	\\ 
EAGB14	&	6252667063729633024	&	7.381	&	0.040	&	4886	&	70	& 158	&	8	&	0.675	&	0.050	\\ 
EAGB15	&	6252665277022961664	&	10.768	&	0.226	&	4938	&	71	& 151	&	8	&	0.914	&	0.070	\\ 
EAGB18	&	6252666445252211072	&	10.166	&	0.272	&	5054	&	79	& 123	&	7	&	0.649	&	0.053	\\ 
RGB10  & 6252664933425563392 & 3.288 & 0.039 & 4615 &	61 & 321 & 18 & 0.747 & 0.054 \\
RGB11  & 6252663765194440064 & 3.092 & 0.043 & 4639 &	64 & 316 & 17 & 0.681 & 0.088 \\
RGB13  & 6252666548333609472 & 3.470 & 0.160 & 4594 &	70 & 308 & 18 & 0.769 & 0.071 \\
RGB14  & 6252665792419450240 & 3.525 & 0.078 & 4594 &	60 & 307 & 17 & 0.778 & 0.059 \\
RGB15  & 6252665861138804096 & 3.262 & 0.253 & 4587 &	60 & 294 & 16 & 0.694 & 0.074 \\
RGB17  & 6252663936993546368 & 3.393 & 0.117 & 4635 &	61 & 277 & 15 & 0.655 & 0.052 \\
RGB18  & 6252664967785297792 & 3.867 & 0.039 & 4636 &	61 & 275 & 15 & 0.741 & 0.054 \\
RGB19  & 6252666995010156928 & 3.972 & 0.117 & 4638 &	63 & 252 & 14 & 0.698 & 0.055 \\
RGB21  & 6252665654980421504 & 4.472 & 0.039 & 4673 &	69 & 235 & 13 & 0.711 & 0.055 \\
RGB22  & 6252666582693327232 & 4.622 & 0.437 & 4623 &	61 & 223 & 12 & 0.726 & 0.086 \\
RGB26  & 6252666823211497728 & 5.390 & 0.039 & 4677 &	63 & 198 & 11 & 0.720 & 0.053 \\
RGB27  & 6252666548333588096 & 6.653 & 0.121 & 4708 &	64 & 194 & 10 & 0.852 & 0.064 \\
RGB32  & 6252678539882310016 & 7.102 & 0.117 & 4751 &	65 & 168 & 9 & 0.765 & 0.057 \\
RGB33  & 6252664929126427392 & 7.200 & 0.086 & 4786 &	66 & 165 & 9 & 0.742 & 0.054 \\
RGB34  & 6252667269888061824 & 7.169 & 0.143 & 4756 &	65 & 161 & 9 & 0.737 & 0.056 \\
RGB37  & 6252665448821987456 & 7.393 & 0.117 & 4736 &	67 & 155 & 8 & 0.742 & 0.056 \\
RGB38  & 6252665036504780672 & 7.839 & 0.195 & 4723 &	65 & 151 & 8 & 0.775 & 0.060 \\
RGB39  & 6252665620620718336 & 8.110 & 0.120 & 4812 &	84 & 145 & 8 & 0.719 & 0.062 \\
RGB40  & 6252666479614120448 & 7.920 & 0.680 & 4764 &	72 & 141 & 8 & 0.707 & 0.081 \\
RGB48  & 6252663902633396224 & 12.271 & 0.363 & 4874 &	81 & 105 & 6 & 0.752 & 0.065 \\
\enddata
\end{deluxetable*}

\section{Inferring seismic masses}
\label{sect:scaling_rel}

We calculated stellar masses using asteroseismic scaling relations which describe how various parameters relate to the solar values \citep{Kjeldsen-Bedding-1995,Kallinger-2010,Huber2011}.
From the four standard seismic mass equations which can be derived \citep{Miglio-2012}, we chose the following:
\begin{equation}
     \left( \frac{M}{M_{\odot}} \right) \simeq \left( \frac{v_{\mathrm{max}}}{\nu_{\mathrm{max},\odot}} \right) \left( \frac{L}{L_{\odot}} \right) \left( \frac{T_{\mathrm{eff}}}{T_{\mathrm{eff},\odot}} \right)^{-7/2}, 
     \label{eq:scaling}
\end{equation}
where the corresponding values of the Sun are $\nu_{\mathrm{max},\odot} = 3090\pm30\,\mu$Hz \citep{Huber2011}, $T_{\mathrm{eff},\odot} = 5772$\,K \citep{Prsa-2016}.
\cite{Maddy2022} already showed that this representation, which avoids any $\Delta\nu$ dependence, is the most accurate for our purposes due to having the smallest uncertainties and the best agreement with previous studies. The issues with $\Delta\nu$ stem from the fact that the K2 time-series are not long enough to fully resolve the large separation pattern for stars with $\nu_{\rm max}\lesssim 50\,\mu$Hz \citep{Hekker2011,Stello2017}. Using the equation above also allows us to directly compare our results with previous ones on K2 globular clusters \citep{Tailo2022, Maddy2022, Maddy2024, Maddy2025}. 

Considering the mode lifetimes that we can reach in this cluster, the K2 light curves may yield only partial presence of excited modes. 
To assess the impact of short data, \cite{Stello-2022} compared one or two sector-long TESS light curves to the original \textit{Kepler} observations of red giants in the \textit{Kepler} field. 
By selecting stars with $\nu_{\rm max}<15\,\mu$Hz and correcting for the K2 campaigns being three times longer than a TESS sector, we estimate a systematic uncertainty of $\sim$8-9\% that our results may carry. Similarly, comparing quarter-to-quarter variations (three-month observing windows) of \textit{Kepler} red giants, \cite{Sreenivas2024} found that typical scatter in $\nu_{\rm max}$ ranges 2-5\%.
However, these values apply to individual stars. By studying more stars at the same evolutionary stages, we can infer ensemble averages for which uncertainties decrease with increasing sample size \citep{Stello-2026}.

Deviations from Eq.~\ref{eq:scaling} for more evolved stars has been reported several times (see \citealt{Hekker-2020} and references therein), possibly due to stellar structures increasingly differing from the Sun. This could introduce systematics into the calculated masses, which are usually treated by obtaining correction factors ($f_{\nu{\mathrm{max}}},\, f_{\Delta\nu}$) from models and placing them in the scaling relations. This is more prominent in $\Delta\nu$, as $f_{\Delta\nu}$ is theoretically motivated, whereas no robust theoretical prediction exists for $\nu_{\rm max}$ \citep{Pinsonneault2018}. Consequently, the inferred $f_{\nu \rm max}$ may also include systematics originating from the $\Delta\nu$ scaling relation or the adopted $T_{\rm eff}$ scale rather than intrinsic deviations in the $\nu_{\rm max}$ relation itself \citep{Pinsonneault-2025}. Moreover, \cite{Ash-2025} showed that using $f_{\nu \rm max}$ calibrated from stellar radii may result in overcorrected masses. For comparison, previous studies of K2 globular clusters found that RGB seismic masses generally agree with stellar evolution models without corrections \citep{Maddy2022,Maddy2024}. 
A possible metallicity dependence of $\nu_{\rm max}$ has also been proposed \citep{Ash-2025,Lindsay2026}, although such a trend may arise from systematic offsets in spectroscopic measurements \citep{Li2022}. Based on many metal-poor ([M/H]$<-1$) APO-K2 stars, for which asteroseismic masses exceed independent astrophysical estimates, \citet{Schonhut-Stasik2024} argued that the discrepancy may originate from offsets in the adopted temperature scale. Similarly, \citet{Yaguang2024} found that $f_{\nu \rm max}$ remains close to unity and that its positive correlation with metallicity inferred from individual frequency modeling largely disappears when adopting a calibrated $\alpha_{\rm MLT}$--[M/H] relation. 

Regardless of these debated issues, such effects primarily influence the individual seismic masses rather than the inferred mass loss, unless $f_{\nu_{\rm max}}$ differs systematically between RGB and EAGB stars, for which no evidence currently exists. A metallicity-dependent $f_{\nu_{\rm max}}$ could nevertheless affect the mass-loss--metallicity trend. However, \citet{Lindsay2026} found a value of $f_{\nu_{\rm max}}\approx1.08$ across the metallicity range spanned by the K2 globular clusters, which remains nearly constant, causing a negligible impact on the inferred metallicity trend. Therefore, we adopt $f_{\nu_{\rm max}}$ as unity.

Besides the seismic information, the scaling relation also requires the luminosity and $T_{\rm eff}$ of each star. 
$T_{\rm eff}$ is only available for 7 luminous red giants in the cluster from spectroscopic measurements by \cite{Koch2014}, all of them are outside of the brightness limits of our RG sample. Therefore, we calculated photometric temperatures following suggestions by \cite{Maddy2022}, which gave consistent results compared to the spectroscopic sample (for further details, see Appendix~\ref{sect:AppendixC}). For the luminosities, beyond the true distance of the cluster, reliable extinction and bolometric correction is necessary as well.

\subsection{Extinction}
\label{sect:extinction}

NGC 5897 is located towards the Libra constellation at higher galactic latitudes compared to most of the globular clusters. The Harris catalog lists a relatively small, $E(B-V)=0.08$\,mag reddening. 
We confirm this value by comparing it with the 3D Bayestar catalog \citep{Bayestar-2019} at the distance of this cluster, showing negligible scatter for the mass calculations. For the extinction coefficient, we assumed $R_V=3.1$ in $A_V=R_VE(B-V)$.

\subsection{Effective temperature}
\label{sect:teff}

The most reliable method to obtain photometric temperature for evolved stars operates with \mbox{\textit{V--K}} color indices \citep{Campbell2017}. We queried 2MASS $K$ magnitudes for our stars, and found that all with detected power excess have valid measurements in the 2MASS catalog 
\citep[quality flag $A$,][]{Skrutskie-2006}. Then, we followed the $T_{\mathrm{eff}}$--color--metallicity relation adopted from \cite{Gonzalez2009} using Equation (10) and coefficients from Table 5 (`Giant stars'). For the required dereddened colors, we converted the $E(B-V)$ reddening of the cluster to $E(V-K)$ based on Equation (8) from \cite{Fitzpatrick2007}. 

\subsection{Luminosity and bolometric correction}
\label{sect:lum}

The luminosities can be obtained in the following way:
\begin{equation}
    \log_{10}(\mathrm{L}/\mathrm{L_{\odot}}) = -0.4 [V_0-(m-M)_0+\mathrm{BC}-M_{bol,\odot}]
\end{equation} 
where $V_0$ is the extinction-corrected visual brightness ($V_0=V-A_V$) and $(m-M)_0=\mu=15.49\pm0.04$\,mag is the true distance modulus (adopted from \citealt{Baumgardt-2021}). BC is the bolometric correction for the $V$ filter, for which we used Equation (18) for giant stars from \cite{Alonso1999}. This also requires the already calculated
$T_{\rm eff}$ and metallicity of the stars. $M_{bol,\odot}=4.74$ is the bolometric magnitude of the Sun \citep{Mamajek2015a}. 

\section{Stellar mass results}
\label{sect:masses}

\begin{figure}
\includegraphics[width=0.95\columnwidth]{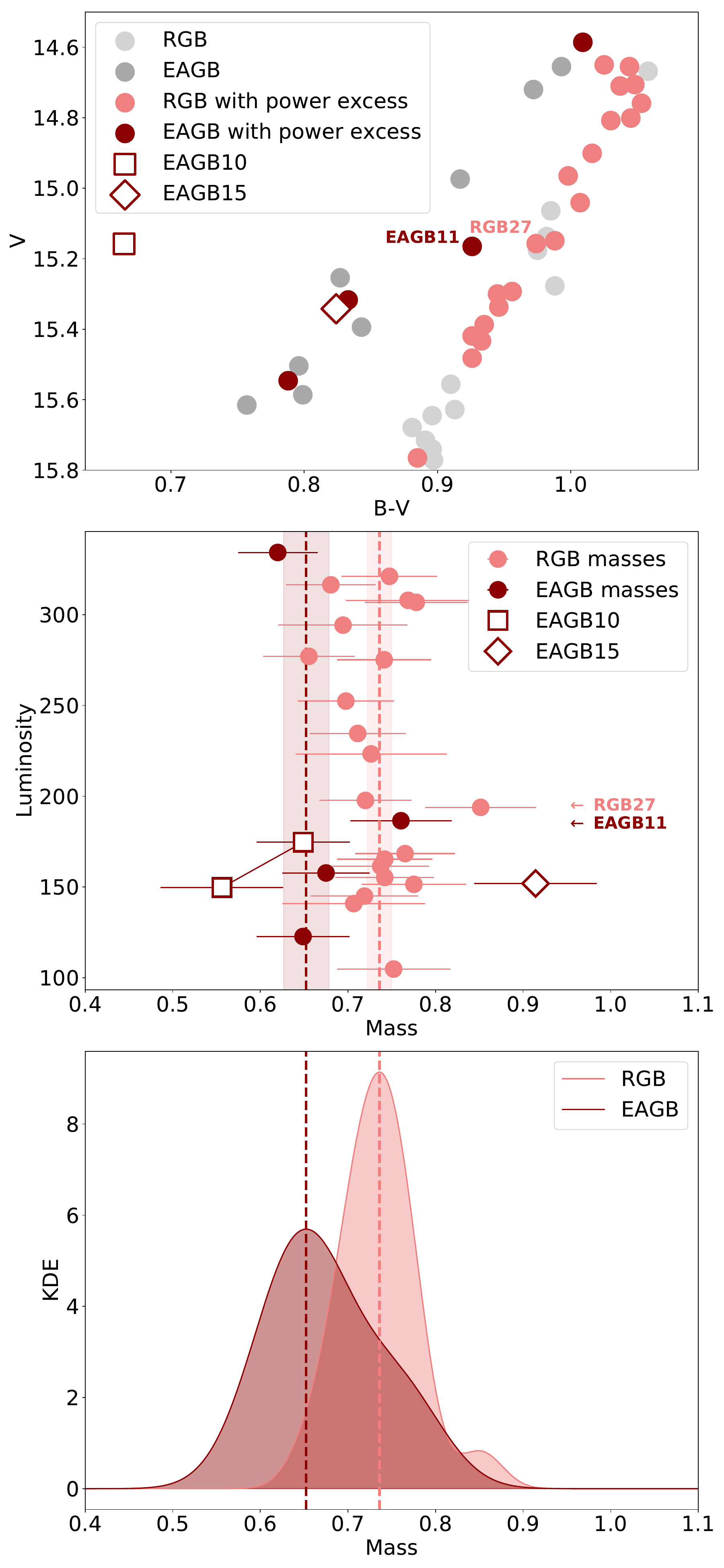}
\caption{Seismic masses of RGB and EAGB stellar populations in NGC 5897. \textbf{Top panel:} EAGB and RGB stars with (colored points) and without (gray points) a successful $\nu_{\rm max}$ detection. 
EAGB10 (square) has an unusual CMD position and is possibly a binary star. \textbf{Middle panel:} Luminosity as function of the individual mass results. Vertical shaded regions in the middle panel indicate the calculated error bars around the KDE peaks from the bottom panel. EAGB15 (diamond) is an overmassive star, presumably a product of binary mass transfer or a stellar merger. 
Significant mass outliers EAGB11 and RGB27 are also indicated. We marked the lower luminosity mass result of EAGB10 too; see the text for details. \textbf{Bottom panel:} Probability distributions of masses of RGB and EAGB stars, using stars with filled markers in the above plots. Vertical dashed lines correspond to the KDE peaks.
\label{fig:mass_results}}
\end{figure}

Figure \ref{fig:mass_results} shows the mass distribution of the RGB and EAGB stars in NGC 5897. The top panel displays the RG sample of the cluster from Figure \ref{fig:BV_CMD}, highlighting stars with detectable oscillations. In the middle panel, the individual mass results and their propagated error bars can be seen, while the dashed lines and shaded regions correspond to the average RGB and EAGB masses and their uncertainties based on the bottom panel from the Kernel Density Estimation (KDE) plot for each evolutionary stage. KDEs were already used in previous studies to determine the characteristic masses of RGB, HB and EAGB stars in K2 clusters \citep{Maddy2022,Maddy2024,Maddy2025}. For consistency, we followed the method in those works, adopting the standard error of the mean as the uncertainty on the average mass estimates.
For RGB stars, we got a very well defined average mass of $\overline{M}_{\rm RGB}=0.74\pm0.01\,M_{\odot}$. Due to the small number of EAGB stars with reliable oscillations, it is harder to constrain the average EAGB mass, although they clearly prefer lower values compared to the RGBs. 
The mode of the KDE distribution gives $\overline{M}_{\rm EAGB}=0.65\pm\,0.03\,M_{\odot}$. 

When calculating the EAGB mass distribution, we excluded two stars: EAGB10 and EAGB15, as possible binaries. Based on the unusual HRD position of EAGB10 (see the top panel of Figure~\ref{fig:mass_results} and Appendix~\ref{sect:AppendixA}), we propose a luminosity contribution of a putative companion to be 15--25 L$_\odot$. After correcting the stellar luminosity of EAGB10 for this effect, the resulting mass can be seen in the middle panel of Fig.~\ref{fig:mass_results} as the lower-mass solution for this star. The position of EAGB15 in the CMD aligns well with the EAGBs of the cluster, therefore we suggest that the companion was slightly more massive, and previously transferred mass to the AGB star. This means that it is already a white dwarf and does not contribute significantly to the luminosity budget. The only significant RGB outlier, RGB27, seems to be analogous to EAGB15. As there are no signs of any offsets in its physical parameters, this overmassive star may have experienced an earlier mass transfer from a more massive companion. EAGB11 seems to be an RGB mass impostor, however, its color is consistently shifted slightly to the red in all different CMDs (see Appendix~\ref{sect:AppendixA}), yielding an ambiguous position between the RGB and EAGB branches that can cause its elevated mass.
More information on the individual EAGB stars can be found in Appendix~\ref{sect:AppendixD}. The derived physical parameters are listed in Table \ref{tab:results}.

\section{Summary and Discussion}
\label{sect:conclusions}

\begin{figure}
\includegraphics[width=1.0\columnwidth]{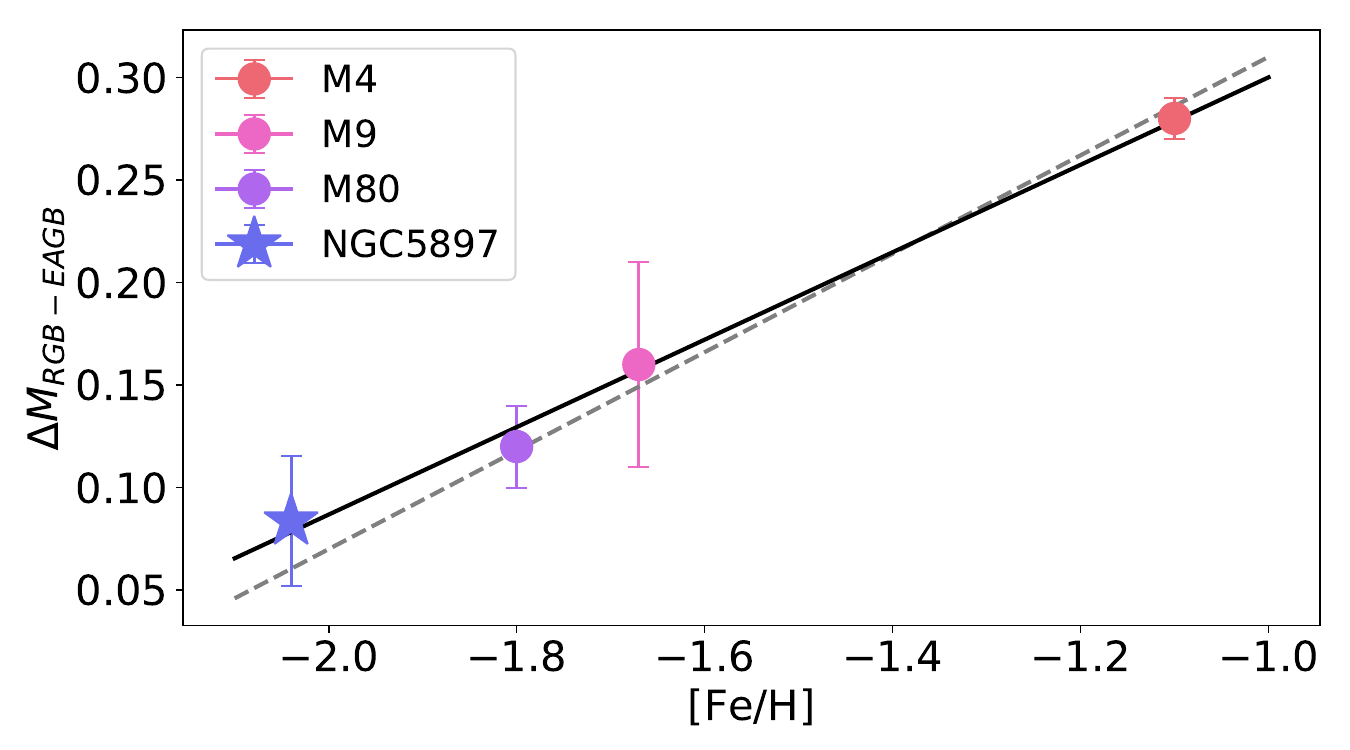}
\caption{Integrated mass loss measurements between the RGB and EAGB phases in Type I K2 globular clusters \citep{Maddy2025}, covering a wide range of metallicities in the subsolar regime. The dashed gray line is the previously fitted mass-loss--metallicity relation by \cite{Maddy2025} for M4 \citep{Maddy2022}, M9 \citep{Maddy2024}, and M80 \citep{Maddy2024}. The blue star represents our result on NGC 5897, and the solid black line is our updated fit for the trend.
\label{fig:mass_loss}}
\end{figure}

In this paper, we report a detection of 26 solar-like oscillators in the globular cluster NGC 5897 for the first time, using \textit{Kepler} K2 observations from Campaign 15. This is the fifth K2 cluster where red giants have been seismically analyzed in this way. We derived the frequency of maximum power excess, $\nu_{\rm max}$, for 20 evolved RGB and 6 early-AGB stars, and determined their masses through the asteroseismic scaling relation in Eq.~\ref{eq:scaling}. We characterized the average RGB and EAGB masses in this cluster, which were found to be $\overline{M}_{\rm RGB}=0.74\pm0.01\,M_{\odot}$ and $\overline{M}_{\rm EAGB}=0.65\pm 0.03\,M_{\odot}$, respectively. Based on their inferred probability distributions, we compute the integrated mass loss between the RGB and EAGB phases: $\Delta M_{\rm RGB-EAGB}=0.08\pm 0.03\,M_\odot$. The uncertainty 
was derived as the quadratic sum of the RGB mass and EAGB mass estimate uncertainties \citep[see earlier studies, e.g.][]{Maddy2022,Maddy2024,Maddy2025}. 

Previously investigated K2 globular clusters with a narrow intrinsic metallicity spread (such as M4, M9 and M80) offer a metallicity--mass-loss trend, with more metal-poor clusters exhibiting less mass loss between the RGB and the EAGB phases \citep{Maddy2025}. We show that our integrated mass loss value for NGC 5897 supports this trend and provides the following updated relation (see also Figure \ref{fig:mass_loss}):
\begin{equation}
\Delta M_{\rm RGB-EAGB} = 0.21\,\rm [Fe/H] + 0.51    
\end{equation} 

Although our work probes the metal-poor regime of the relation, increasing the sample size would be essential, with a particular regard for metal-rich globular clusters. Future facilities capable of delivering high-resolution time-series measurements, such as \textit{Roman} or the proposed HAYDN mission \citep{Miglio2021}, combined with the tools of asteroseismology, may provide new opportunities to further advance the theoretical framework of mass-loss in low-mass stars.

\begin{acknowledgements}
We thank the anonymous referee for their valuable suggestions. This research was supported by the `SeismoLab' KKP-137523 \'Elvonal grant of the Hungarian Research, Development and Innovation Office (NKFIH) and by the LP2025-14/2025 Lendület grant of the Hungarian Academy of Sciences. A.P. acknowledges the K-138962 NKFIH grant. This paper includes data collected by the \textit{Kepler} mission and obtained from the MAST data archive at the Space Telescope Science Institute (STScI). Funding for the Kepler mission is provided by the NASA Science Mission Directorate. STScI is operated by the Association of Universities for Research in Astronomy, Inc., under NASA contract NAS 5–26555. This work has made use of data from the European Space Agency (ESA) mission \textit{Gaia} (\url{https://www.cosmos.esa.int/gaia}), processed by the \textit{Gaia} Data Processing and Analysis Consortium (DPAC, \url{https://www.cosmos.esa.int/web/gaia/dpac/consortium}). Funding for the DPAC has been provided by national institutions, in particular the institutions participating in the \textit{Gaia} Multilateral Agreement. This research made use of NASA’s Astrophysics Data System Bibliographic Services, as well as of the SIMBAD and VizieR databases operated at CDS, Strasbourg, France.

\end{acknowledgements}

\facilities{\textit{Kepler} K2 \citep{Howell-2014,K2-C15}, \textit{Gaia} \citep{Gaia-2016}}
\software{lightkurve \citep{lightkurve2018}, 
        K2SC \citep{Aigrain2016},
        fitsh \citep{PAl2012},
        pyMON \citep{Maddy2025}.
          }

\bibliography{references}{}
\bibliographystyle{aasjournal}

\appendix 
\section{Comparison of color-magnitude diagrams using different photometries}
\label{sect:AppendixA}

\begin{figure*}[t!]
\includegraphics[width=0.33\textwidth]{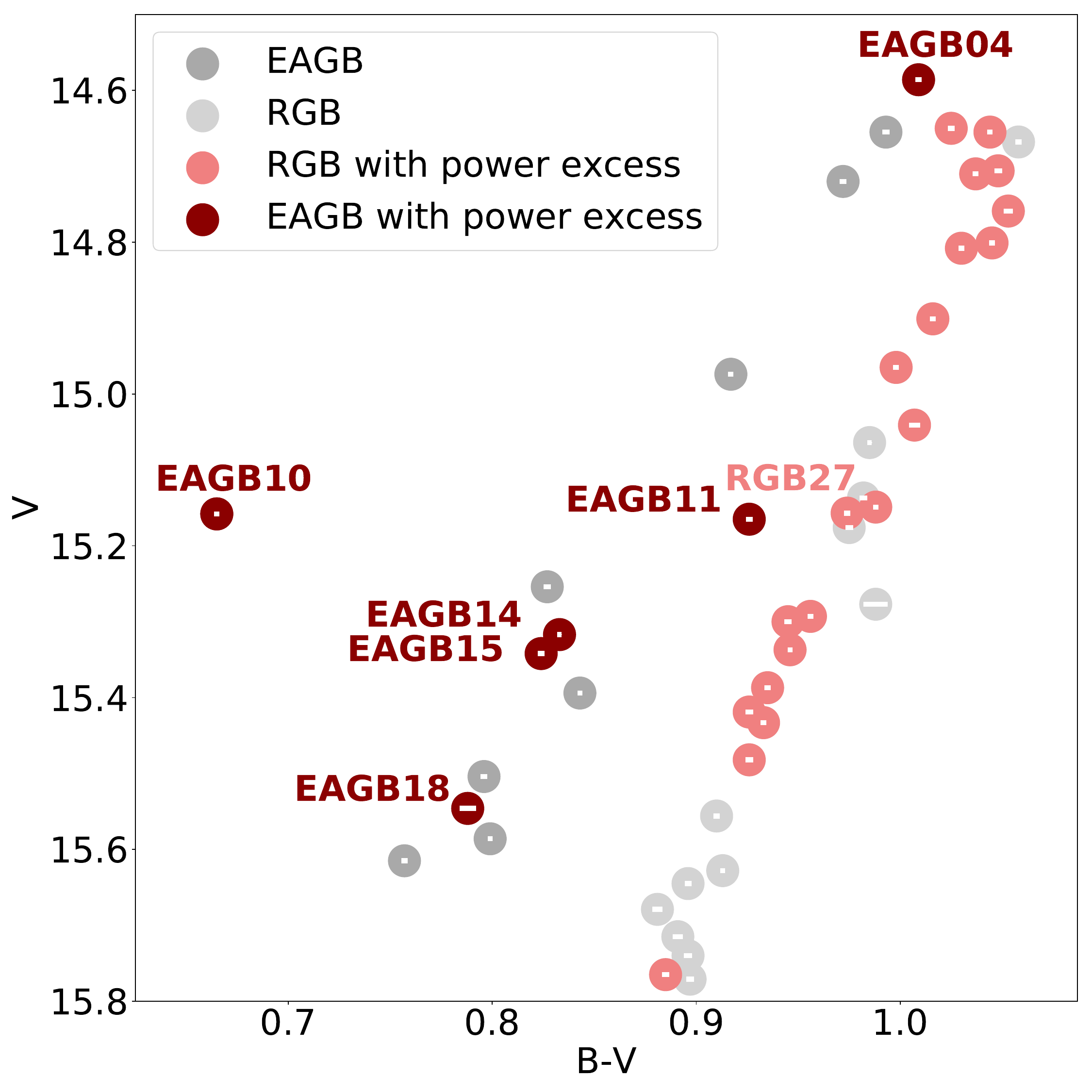}
\includegraphics[width=0.33\textwidth]{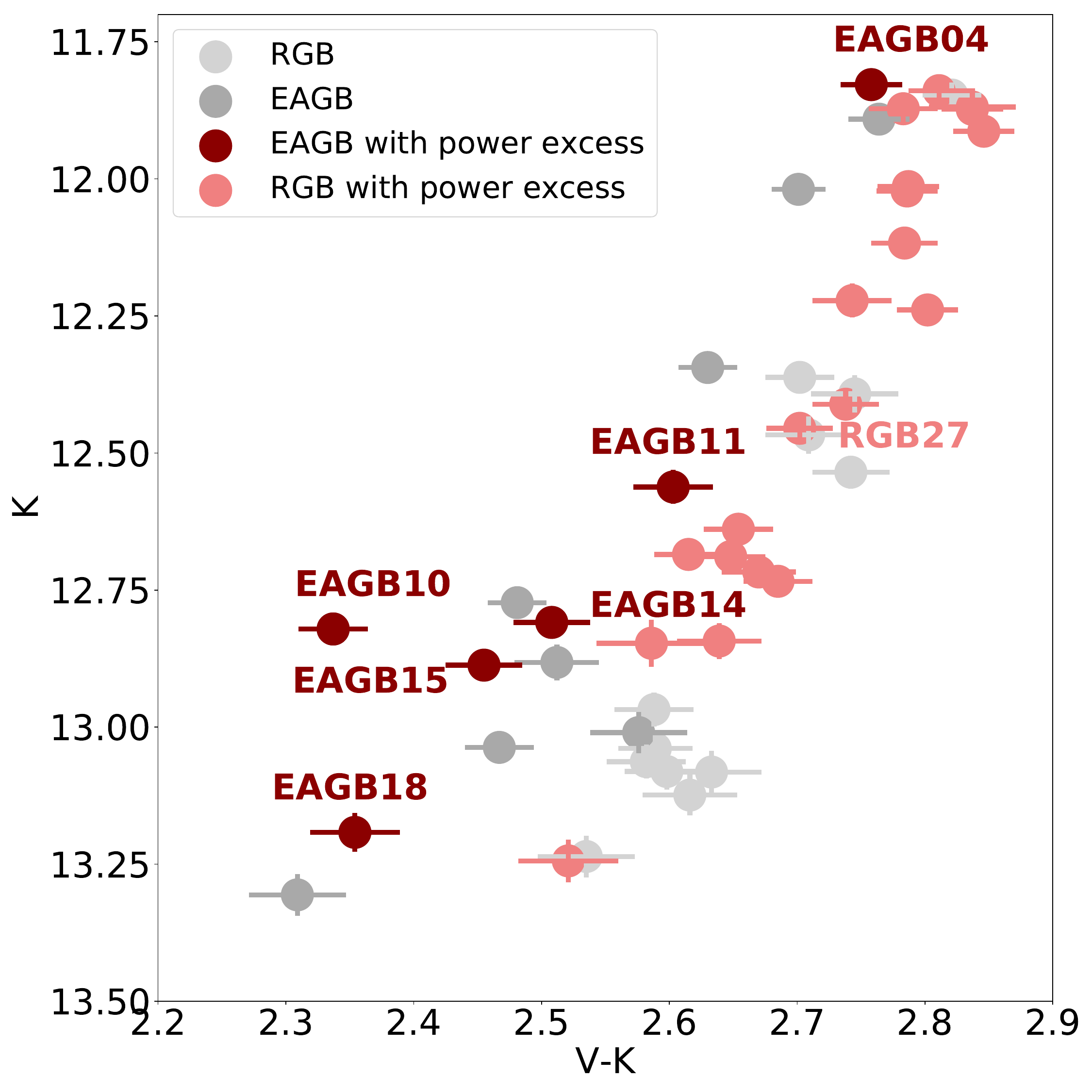}
\includegraphics[width=0.33\textwidth]{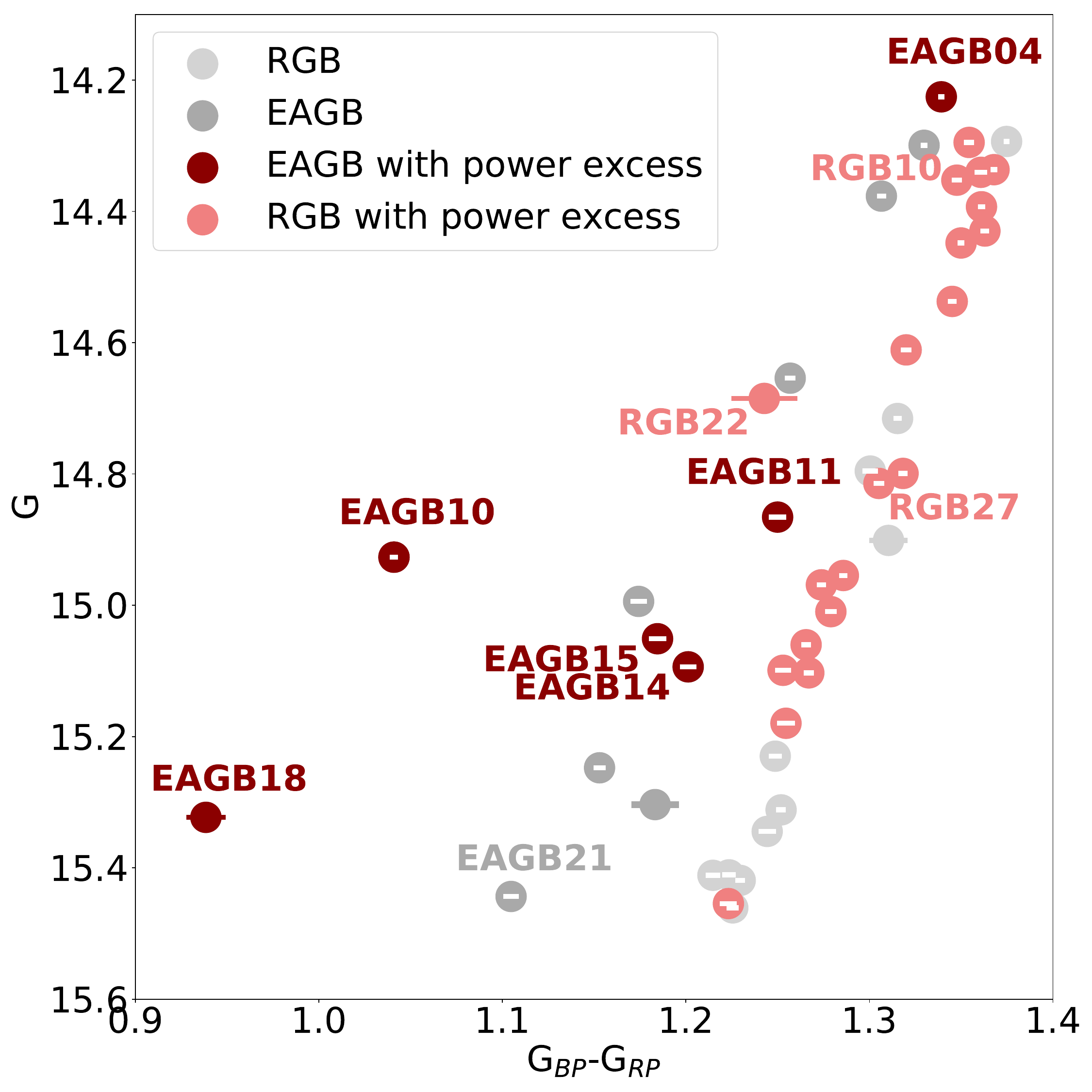}
\caption{Comparison of different color-magnitude diagrams of the relevant red giant sample of NGC 5897 by using $B$ $V$, and $K$ photometry from \cite{Stetson2019} and \cite{Skrutskie-2006}, and \textit{Gaia} $G$, $G_{\rm BP}$, $G_{\rm RP}$ colors by \cite{Gaia-EDR3-2021}. When the photometric errorbar of the data point smaller than the size of the point, the errorbar is shown by white inside the points. Our successful detection of solar-like oscillators are in color.
\label{fig:CMD_detection}}
\end{figure*}

In order to evaluate our classification of the targets into RGB and EAGB stages, we compared different color--magnitude diagrams (CMDs) constructed from various passbands and photometric datasets. The left panel of Fig.~\ref{fig:CMD_detection} shows the (\textit{B--V}) CMD by \citet{Stetson2019}, which is also presented in the top panel of Figure \ref{fig:mass_results} in the main text. The photometric uncertainties are smaller than the symbol sizes and are indicated by white lines within each circle. The middle panel displays the (\textit{V--K}) colors of these stars, based on the Stetson $V$ photometry and the near-infrared $K$-band data from the 2MASS survey \citep{Skrutskie-2006}. Although the error bars are larger, the RGB and EAGB distribution remains consistent with the previous panel. The right panel presents the \textit{Gaia} DR3 CMD using $G_{\rm BP},\,G_{\rm RP}$ and $G$, providing colors that are completely independent from the previous CMDs. Despite the seemingly small photometric uncertainties, \textit{Gaia} measurements can be significantly affected by crowding and contamination, to which the ($G_{\rm BP}-G_{\rm RP}$) color is particularly sensitive. An indicator known as the BP and RP flux excess factor ($C$, queried as \texttt{phot\_bp\_rp\_excess}) provides the ratio between the total flux in BP and RP and the $G$-band flux, and is used to test whether the $G_{\rm BP}$ and $G_{\rm RP}$ photometry are consistent with the $G$-band measurement \citep{Evans2018}. To further refine this diagnostic tool, \cite{Riello2021} introduced the $C^*$ corrected color excess factor, which removes the strong color dependence of $C$. Ideally, $C^* \approx 0$; any significant deviation indicates inconsistency in the photometry. 

\begin{figure}
\includegraphics[width=1\columnwidth]{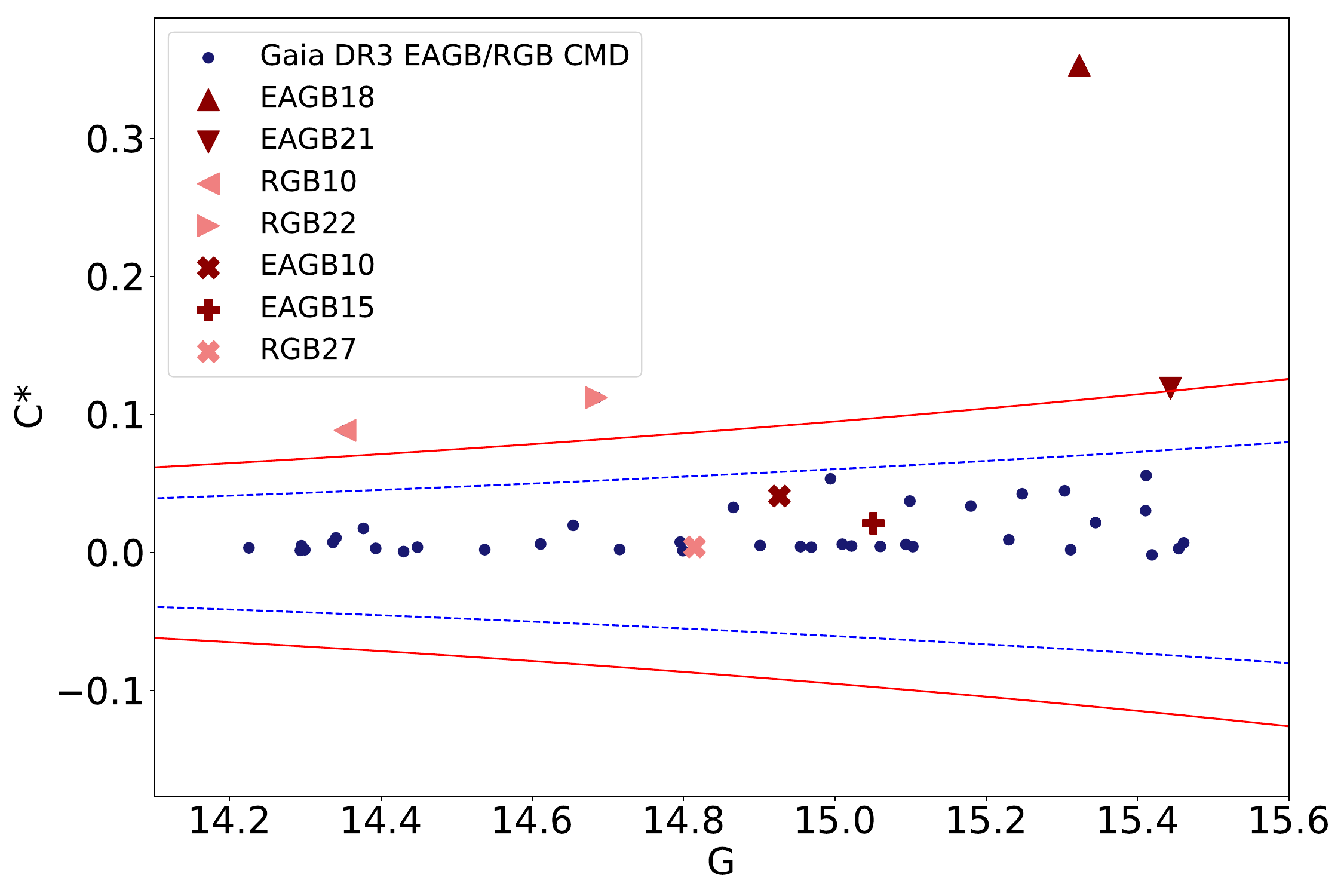}
\caption{Calculated corrected color excesses ($C^*$) based on \cite{Riello2021}, in order to evaluate the \textit{Gaia} DR3 $G_{\rm BP}$ and $G_{\rm RP}$ photometry against $G$. The dark blue points are stars from panel $c$ of Figure~\ref{fig:CMD_detection}. Overmarks in light pink and dark red represent RGB and EAGB stars, respectively. Dashed blue lines corresponds to the $7\sigma$ scatter, while the solid red curves are for $10\sigma$, see more information in Appendix~\ref{sect:AppendixA}. This figure confirms the nature and position of our sample stars, with a particular regard for EAGB10, EAGB11 and RGB27.
\label{fig:corr_color_exc}}
\end{figure}

We computed $C^*$ for all stars shown in panel $c$ of Figure \ref{fig:CMD_detection} and found that 42 out of 46 stars satisfy $C^* < 0.055$ (see Figure~\ref{fig:corr_color_exc}). This threshold corresponds to the $7\sigma$ scatter derived for well-behaved, isolated stellar sources with good quality \textit{Gaia} photometry \citep[see Eq. (18) in][]{Riello2021}. Only four stars -- RGB10, RGB22, EAGB18, and EAGB21 -- lie above this limit, with $C^*$ values of 0.086, 0.112, 0.353 and 0.119, respectively (all corresponding to $>10\sigma$ deviations). This clearly shows that the higher corrected color excesses in RGB22 and EAGB18 likely explain their color offsets, while supporting the interpretation that EAGB10 is a physical color outlier, as it exhibits a typical $C^*$ value within the sample. 
Interestingly, EAGB11 still holds its intermediate position between the RGB and EAGB branches in this CMD as well, which may be eventually responsible for its position as an RGB-mass impostor.

\section{Corrected light curves for EAGB and RGB stars}
\label{sect:AppendixB}

\begin{figure*}[t!]
    \includegraphics[width=1\textwidth]{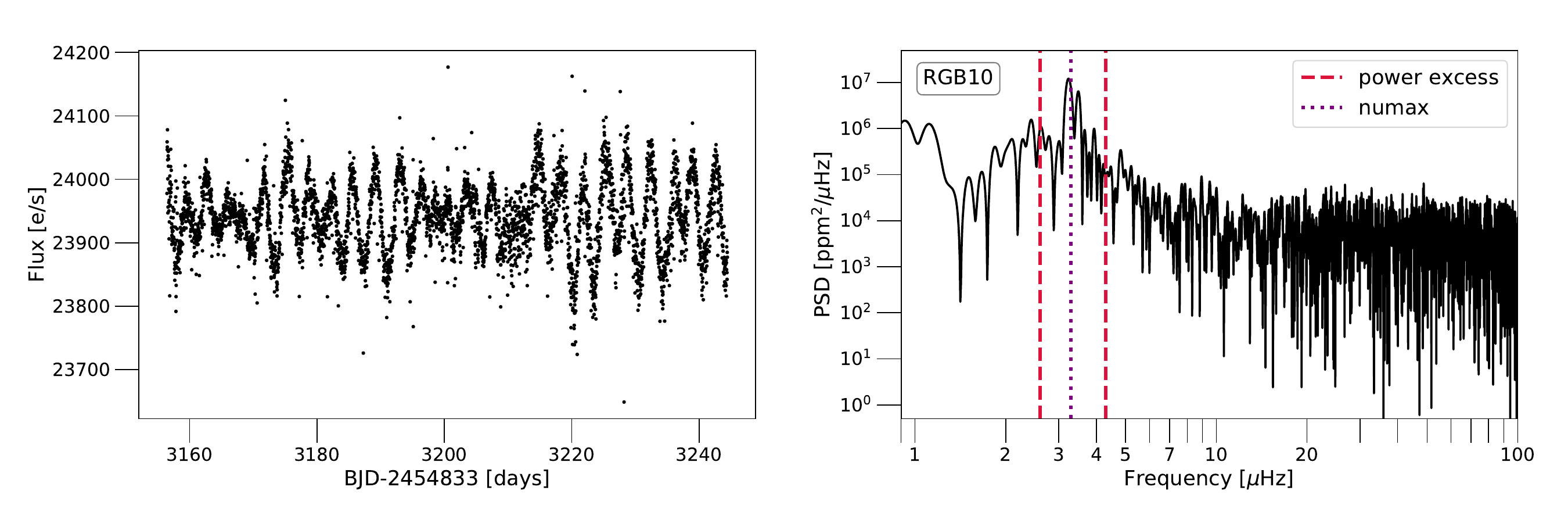}
    \includegraphics[width=1\textwidth]{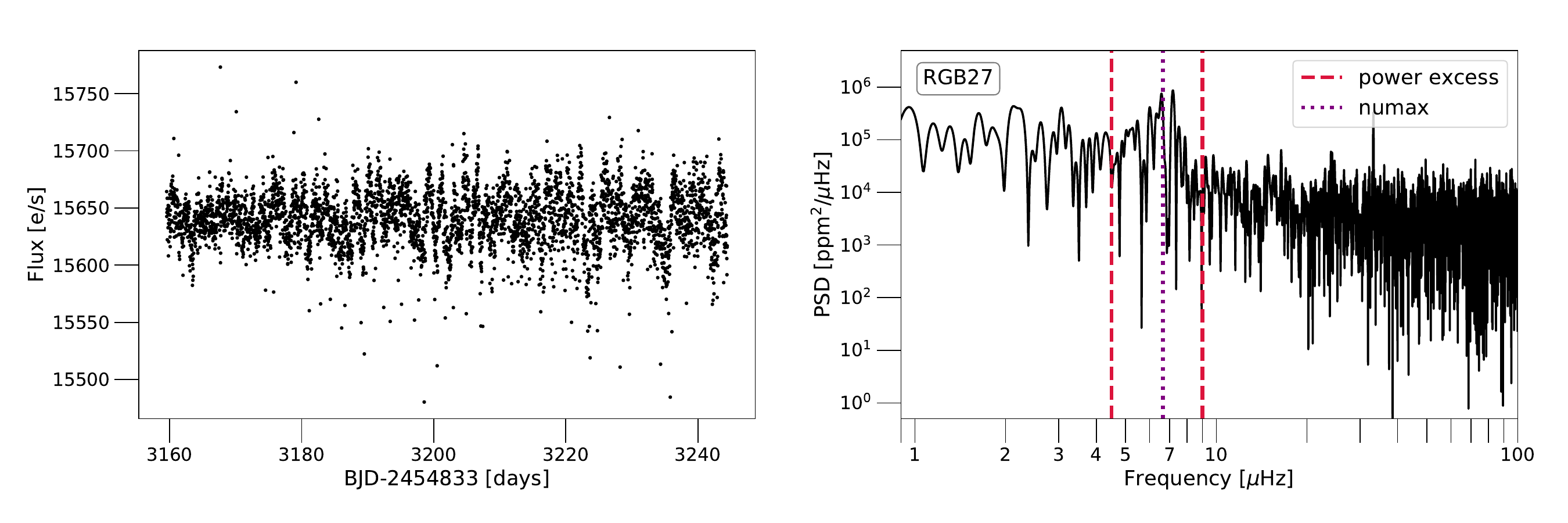}
    \includegraphics[width=1\textwidth]{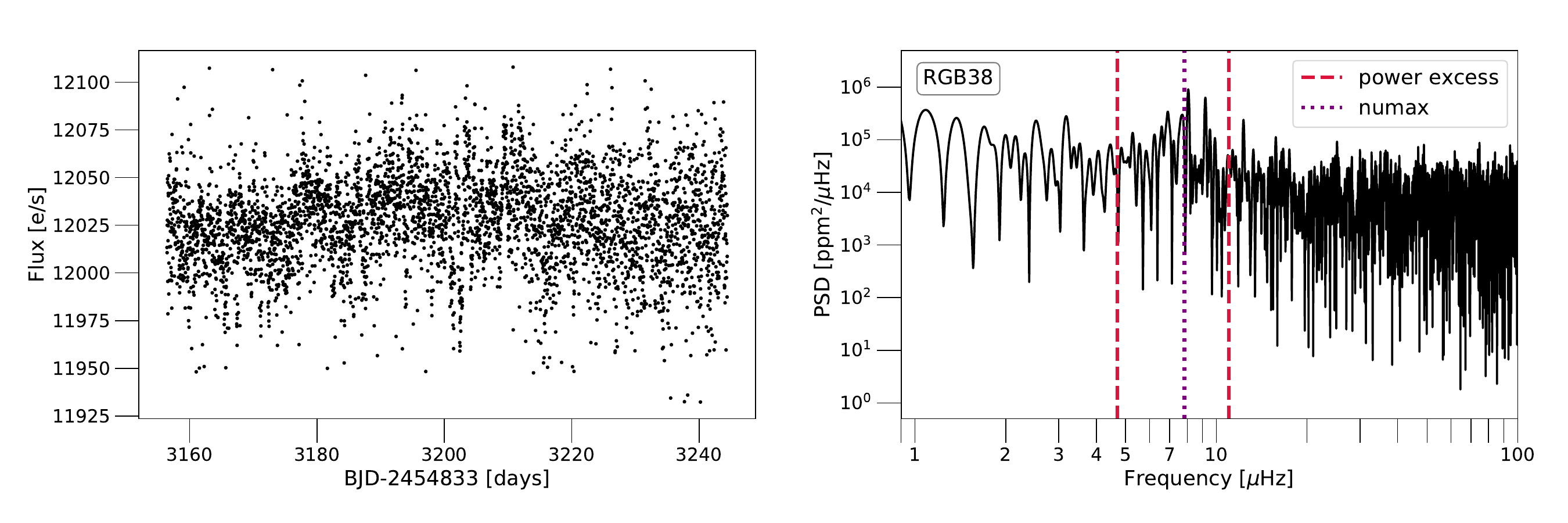}
    \includegraphics[width=1\textwidth]{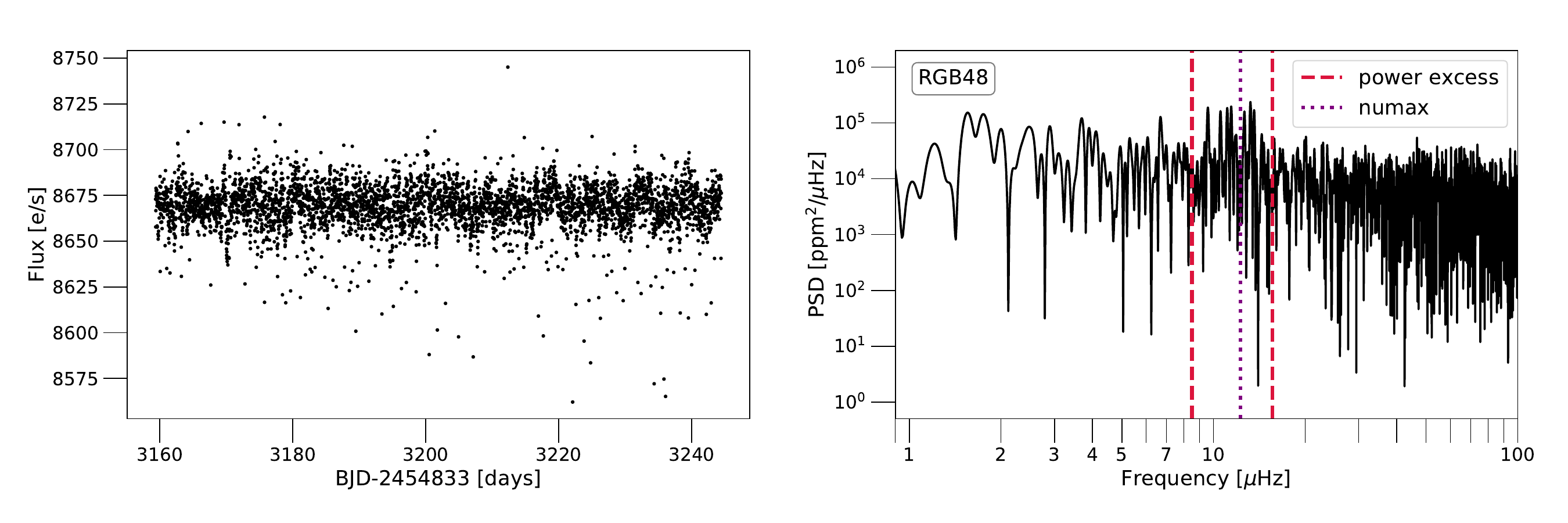}
\caption{Examples for the variety of light curves and power spectra of RGB stars in NGC 5897.}
\label{fig:rgbs}
\end{figure*}

\begin{figure*}[t!]
    \centering
    \includegraphics[width=1\textwidth]{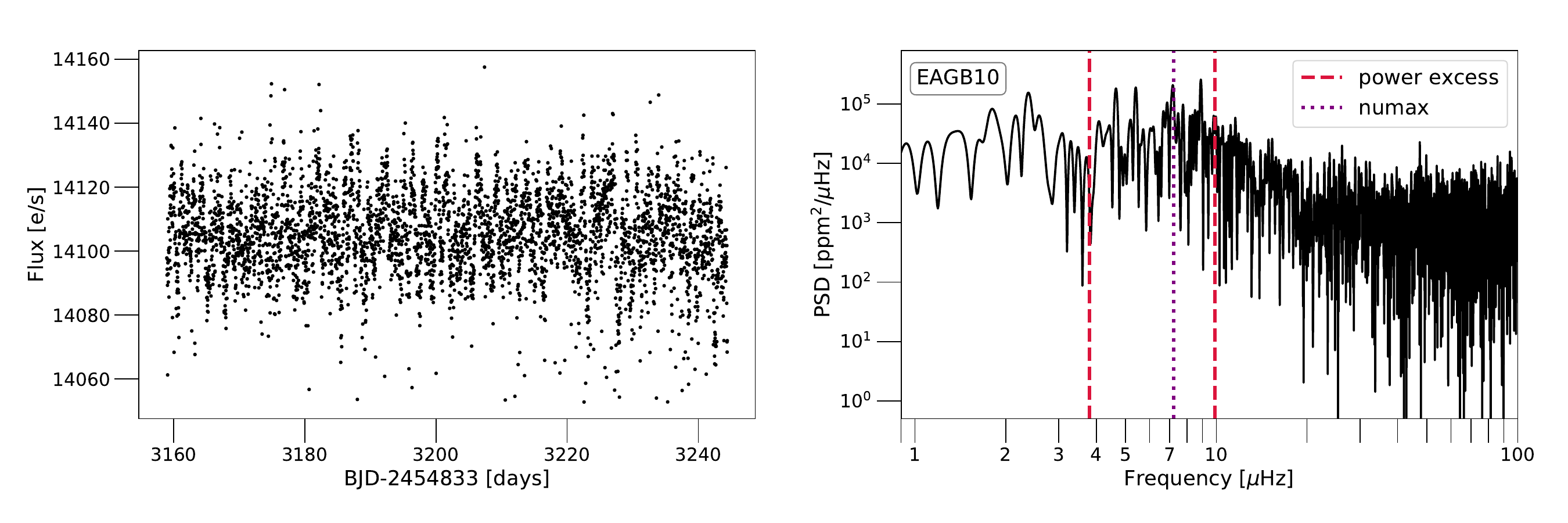}
    \includegraphics[width=1\textwidth]{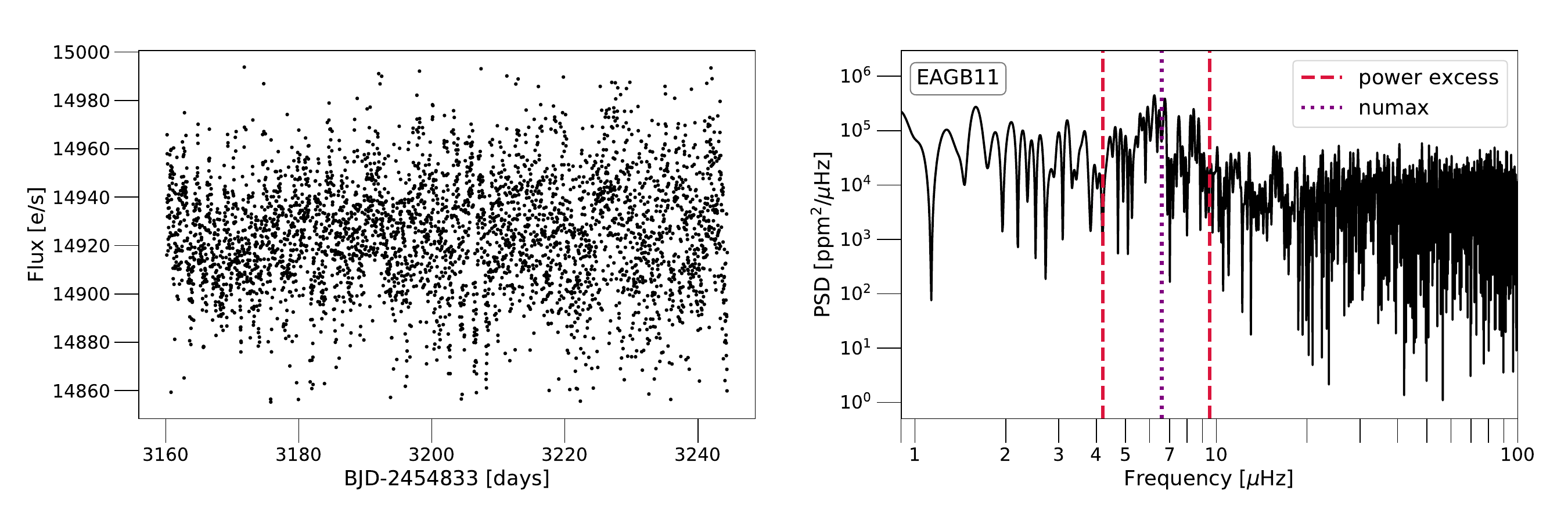}
    \includegraphics[width=1\textwidth]{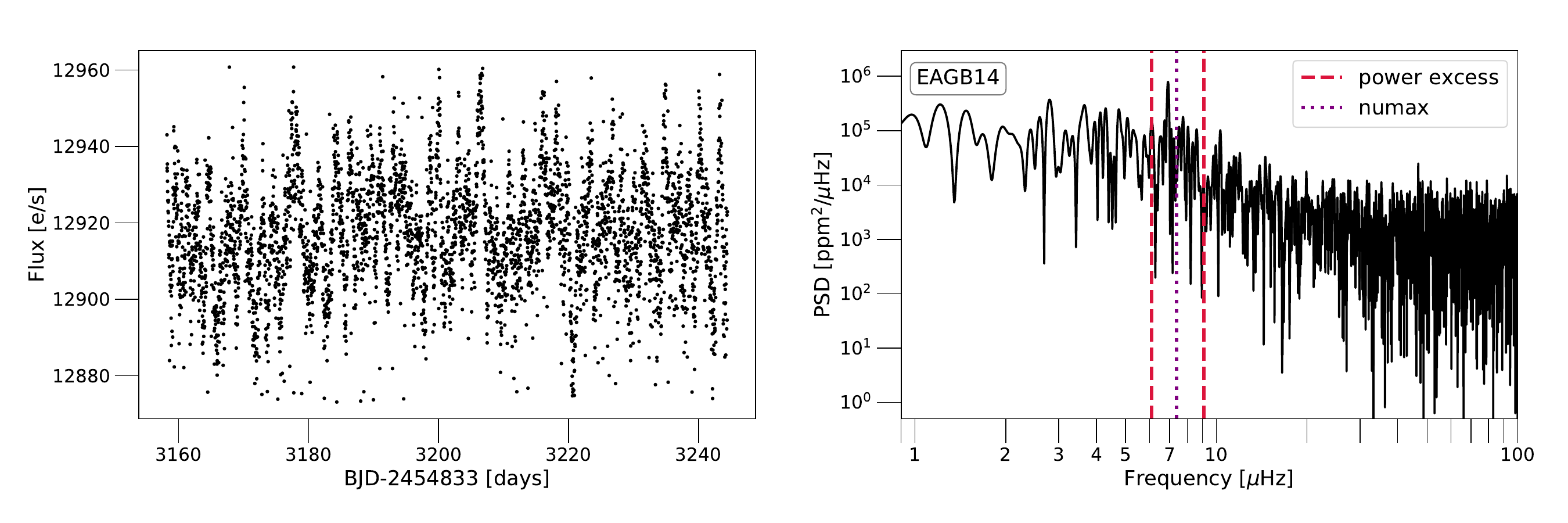}
\caption{As in Figure~\ref{fig:rgbs}, but for EAGB stars in NGC 5897.}
\label{fig:eagbs}
\end{figure*}

As described in Section~\ref{sect:data}, we applied differential aperture photometry to the long-cadence Target Pixel Files of globular cluster NGC 5897, observed in Campaign 15 of the \textit{Kepler} K2 mission. A sample of these light curves, after corrections for instrumental and systematic effects, and containing additional parameters from the \textit{Gaia} DR3 catalog, is shown in Table~\ref{tab:lc}\footnote{Data will be available in its entirety at CDS.}. The time stamps were converted to Baricentric Julian Date (BJD). The differential light curves were shifted to a median flux level determined using the \textit{Gaia} DR3 $G$ magnitudes as a reference, since the bandpasses and magnitude systems of the two missions have been found to correlate well \citep[see][for more details]{Molnar2018a}. We then applied the \textit{Kepler} flux calibration of \cite{Lund2015} to derive the \textit{Kepler} magnitudes. In Figures~\ref{fig:rgbs} and \ref{fig:eagbs}, we show a gallery of light curves and power spectra for some of our RGB and EAGB stars in this cluster.

\begin{deluxetable*}{lccccc}
\tablecaption{Sample table of the photometry of RGB and EAGB stars observed in Campaign 15 by the \textit{Kepler} K2 mission. The table includes the evolutionary stage and identification numbers of the stars, the time stamps of each frames in Baricentric Julian date (BJD\,--\,2\,454\,833), the measured \textit{Kp} brightness and its uncertainty from each frame, plus the corresponding \textit{Gaia} DR3 source ID and mean $G$ magnitude. The entire table is available online at CDS. 
\label{tab:lc}}
\tablehead{
\colhead{ID} & BJD-2454833 [days] & \colhead{Flux [e/s]}  & \colhead{$\sigma_{\rm flux}$ [e/s]} & \colhead{Gaia DR3 source ID} & \colhead{$G$ mean [mag]}}
\startdata
EAGB04	&	3163.739120	&	26941.1	&	26.1	&	6252665788119836672 	&	14.225456	\\ 
EAGB04	&	3163.759552	&	26966.4  	&	25.5	&	6252665788119836672 	&	14.225456    \\ 
EAGB04	&	3163.779985	&	26946.1	&	25.0	&	6252665788119836672 	&	14.225456	\\ 
EAGB04	&	3163.800417	&	26956.4	&	24.8	&   6252665788119836672 	&	14.225456	\\ 
EAGB04	&	3163.820849	&	26959.7	&	24.4	&	6252665788119836672 	&	14.225456	\\ 
\multicolumn{6}{l}{\dots}\\
\enddata
\tablecomments{Table \ref{tab:lc} is published in its entirety in the machine-readable format. A portion is shown here for guidance regarding its form and content.}
\end{deluxetable*}

\section{Comparison of spectroscopic and photometric temperatures}
\label{sect:AppendixC}
\begin{figure}[t!]
\centering
\includegraphics[width=1\columnwidth]{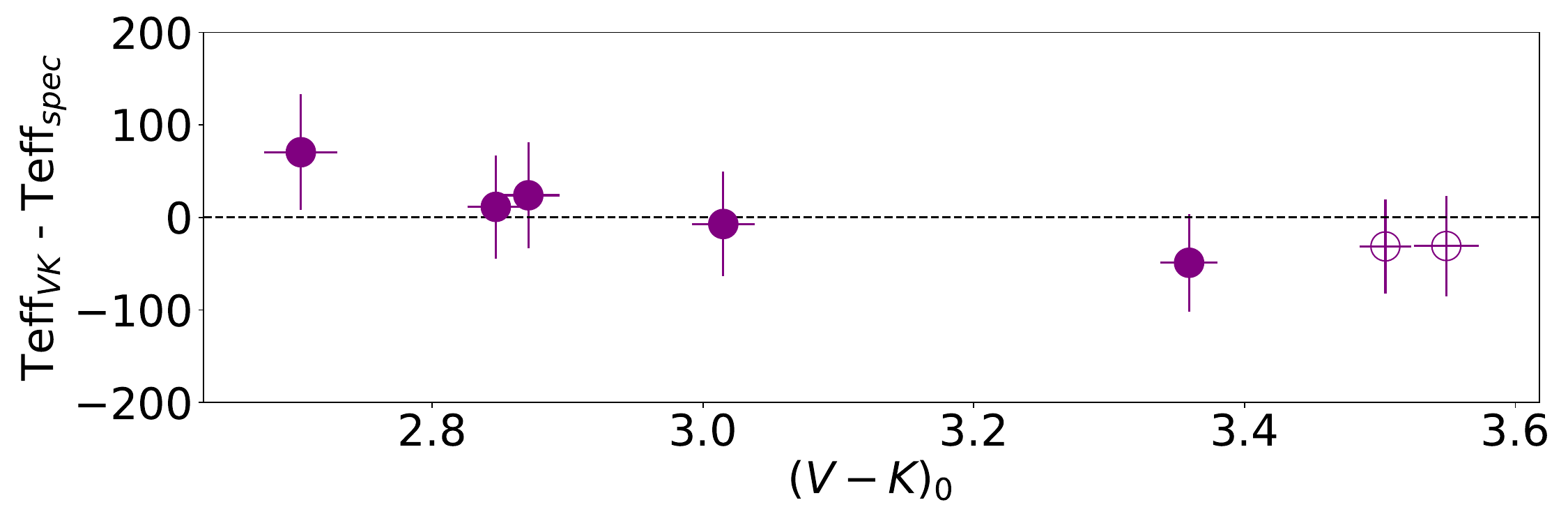}
\caption{Comparison of spectroscopic effective temperatures ($T_{\rm eff_{spec}}$) with calculated photometric temperatures ($T_{\rm eff_{VK}}$) for 7 luminous red giants in NGC 5897 observed by \cite{Koch2014}.}
\label{fig:teff_comp}
\end{figure}

Seismic mass calculations require effective temperatures for each star. Unfortunately, only seven luminous red giants were previously observed spectroscopically by \cite{Koch2014}, and all of them have significantly higher luminosities than the RGB and EAGB stars in our sample. Therefore, we derived photometric temperatures for our targets using $V$ and $K$ magnitudes, the most reliable filter combination for this purpose \citep{Campbell2017,Maddy2022}. To test whether an offset exists between the two methods, we computed the photometric temperatures of these luminous red giants and compared them with their spectroscopic values. Figure \ref{fig:teff_comp} shows the resulting residuals. For the coolest stars, shown with empty circles, the reddening-corrected $(V-K)$ colors lie outside the applicability range of the color--metallicity--$T_{\rm eff}$ relation of \cite{Gonzalez2009}; therefore, this represents a minor extrapolation of the relation. Given the small comparison sample, we did not apply any offset to our photometric temperature estimates.

\section{Statistical and individual analysis of EAGB stars}
\label{sect:AppendixD}

\begin{figure*}[t!]
    \includegraphics[width=1\textwidth]{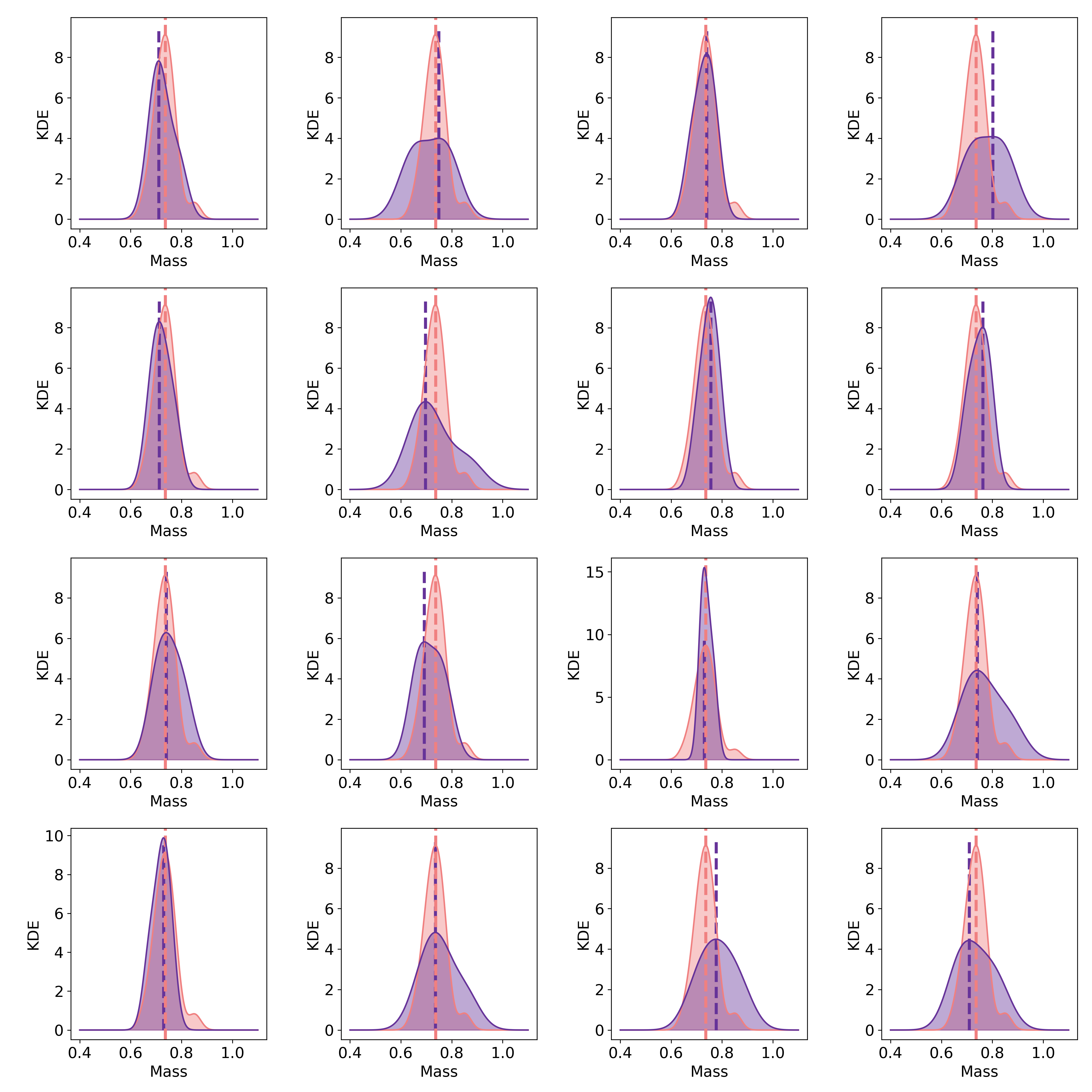}
    \caption{We resampled the original mass distribution of RGB stars (light pink curve) to create new mass distributions of $N=4$ small-sized subsamples (purple curves). The dashed lines represent the peaks of each distribution. After 100,000 random realizations, the purple peak values were used to characterise the uncertainty in the average EAGB mass introduced by small-number statistics. This is an example of 16 random realizations. See Appendix~\ref{sect:AppendixD} for more details.}
    \label{fig:eagb_stats}
\end{figure*}

In our EAGB sample, only four stars could be used to characterise the average EAGB mass of this cluster. Interestingly, such a small sample size of EAGB stars is not unusual among previously studied K2 globular clusters \citep[see][]{Maddy2022,Maddy2025}. While it is worth discussing the individual analysis of each target to ensure their credibility, it is also important to consider how adequately can we capture the inferred uncertainty in $M_{\rm EAGB}$ with such a small sample size.
To assess this impact, we performed the following Monte Carlo simulation. First, we assumed that our larger RGB sample provides a well-defined mass distribution that does not change significantly after evolving to the EAGB. Therefore, the shape of the mass distribution for EAGB stars would be similar for a larger sample size as well. Then, we examined how accurately we could reconstruct the average evolutionary stage mass using only four stars. We randomly drew $N=4$ mass values from the RGB distribution and calculated their corresponding KDEs 100,000 times. Figure~\ref{fig:eagb_stats} shows a few examples of these random realizations, where the light pink curve represents the original RGB mass probability distribution and the purple curves show the newly derived KDEs for the small-size subsamples. Even when sampling only four stars, the peaks of each distribution, indicated by dashed lines, remain close to each other. It can also be noticed that some realizations resemble the overall shape of the observed EAGB mass distribution, which provides additional context behind its asymmetric nature.
After each resampling, a peak was obtained for that realization. The standard deviation of the resulting 100,000 peaks around the true peak mass was found to be equivalent to the central 67\% interval of their probability distribution, which corresponds to an approximate $1\sigma$ uncertainty. This means that the average mass provided by four stars from a population can estimate the true peak mass of the underlying stellar population to this level of precision. This can be translated into an error bar of $\pm0.027\,M_\odot$, which is within the indicated EAGB mass uncertainty of $\pm0.03\,M_\odot$. Therefore, we conclude that the effect of small-number statistics on our final EAGB mass and integrated mass-loss estimates does not dominate over that of other factors in the error budget.

In the following subsections, we discuss further details of the individual frequency analyses of all of our EAGB stars.

\subsection{EAGB04, power excess around $3\,\mu$Hz}

The light curve of EAGB04 shows clear variations driven by a few oscillation modes, while the overall oscillation frequency remains stable throughout the K2 baseline. We determined the $\nu_{max}$ of its power excess to be $2.709\pm0.044\,\mu{\rm Hz}$, resulting in a mass estimate of $m=0.620\pm0.045$\,M$_\odot$. This star falls in the sub--$3\,\mu$Hz regime. \citet{Yu-2020} showed that radii inferred from scaling relations for stars with oscillations below $3\,\mu$Hz tend to increasingly deviate from \textit{Gaia} radii. However, \cite{Ash-2025} directly tested the breakdown of the asteroseismic scaling relations in luminous red giants and found that the $\nu_{\rm max}$-only scaling relation we adopt still appears to hold in this regime. They showed that the inclusion of $\Delta\nu$ in the scaling relations contributes significantly more to the observed breakdown than deviations associated with $\nu_{max}$. For the $\nu_{max}$-only relation, their deviations remain $\lesssim2\%$, corresponding differences only at the second decimal place in our derived stellar masses, well below the uncertainties. Therefore, the relation can be safely extended to this regime.

\subsection{EAGB10, a possible binary}

The position of EAGB10 is curious, as it lies significantly blueward from the rest of the EAGB stars. Unlike some other stars that appear blueward only in the \textit{Gaia} or Johnson data, this star shows the same offset in both CMDs, making color measurement errors highly unlikely (see also Appendix~\ref{sect:AppendixA}). This offset in position could be caused by different effects. It could indicate, for example, a lower-mass stripped giant that have been evolving towards the AGB through the Type II Cepheid instability strip at higher luminosities than the rest of the stars. However, the relatively high mass of the target,  $M=0.65\pm0.05$\,M$_\odot$, makes this scenario unlikely, since Type II Cepheids are expected to be around 0.50--0.55\,M$_\odot$ \citep{smolec-2016,bono-2020}. We compared the position of EAGB10 and the rest of the AGB stars to public HB/AGB tracks computed with the Dartmouth Stellar Evolution Program \citep[DSEP,][]{DSEP-2008}, but found that at the [Fe/H] value of the cluster, the tracks miss the AGB and evolve along higher temperatures, preventing us from drawing clear conclusions.

Another possibility is that the star experiences less interstellar extinction than we calculated, although the differential reddening maps show mostly homogeneous reddening in front of the cluster. But assuming a lower L and $T_{\rm eff}$ together would result in an even higher mass.

We instead propose that the object is a binary, and thus it appears overluminous in the data. Lowering its luminosity by as little as 15--25\,L$_\odot$ also lowers the mass of the star to 0.59--0.55\,M$_\odot$, which would still agree within errorbars with the masses and positions of the rest of the AGB stars in the HRD. This suggests that EAGB10 could be a binary star, with an early RGB star as a companion, which is too faint to be detected via asteroseismology. This assumption is corroborated by the fact that the star has a very high RUWE (Renormalized Unit Weight Error) of 5.28 in \textit{Gaia} DR3. While RUWE is not a direct measurement of unresolved binarity but an indicator of increased deviations from the single-star astrometric solution, it correlates strongly with binarity, with the threshold value value being between 1.15--1.37 at different parts of the sky \citep{Castro-Ginard-2024}. 

\subsection{EAGB11, close match with RGB masses}

EAGB11 shows a prominent power excess in its power spectrum, yielding $\nu_{\rm max}=6.593\pm0.082\,\mu{\rm Hz}$. This corresponds to a stellar mass of $M=0.761\pm0.058\,M_{\odot}$, which is surprisingly in close agreement with the average RGB mass. Due to the small number of characterized EAGB power excesses, it is difficult to assess how strongly this star deviates from the mean EAGB mass. Measuring nearly the same mass as the RGB stars may indicate a very small amount of mass loss, remaining undetectable within precision of the K2 data. On the other hand, a potential misclassification of the star provides an alternative explanation. Interestingly, this star occupies an ambiguous position between the RGB and EAGB branches in all CMDs (see Figure \ref{fig:CMD_detection}). This physically motivated color offset, propagated through the scaling relation, may explain its closer match with the RGB masses rather than with the remaining EAGB population.

\subsection{EAGB14}

The temporal coverage of the K2 data allows us to capture only a snapshot of the oscillation of a given star. This can result in power excesses with unusual shapes, depending on the actual modes excited at the time of the observation. An example was presented by \citet{Molnar-2025} for a bright red giant, where one sector of TESS data exhibited a power excess dominated by a high-amplitude mode that was entirely absent from the power excess observed in another sector four years earlier. In the case of EAGB14, we observe the opposite behavior: the oscillation pattern is characterized by nearly a single excited peak (see Figure~\ref{fig:eagbs}). Although the $\nu_{\rm max}$ uncertainty is measured correctly and consistently, the realistic errorbars may be larger. The fitted power excess was found to be $\nu_{max}=7.381\pm0.040\,\mu{\rm Hz}$, corresponding to a seismic mass of $M=0.675\pm0.050\,M_{\odot}$. This supports the suggestion that the mass loss experienced by stars in this cluster between these two evolutionary phases is relatively small.

\subsection{EAGB15, a more massive star}

EAGB15 shows several low-amplitude peaks within its estimated power excess at $\nu_{max}=10.768\pm0.226\,\mu{\rm Hz}$. This places the star among overmassive objects with a derived mass of $M=0.914\pm0.070\,M_{\odot}$, exceeding the average RGB mass. We propose that it is a binary system in which a slightly more massive companion previously transferred mass to the AGB star and has since evolved into a white dwarf that no longer contributes significantly to the luminosity of the system. Unlike EAGB10, the position of this star in the CMD aligns with the AGB sequence of the cluster, thus any luminosity contribution from a putative companion must be very small. Alternatively, the star could also be a merger product, and thus possibly a descendant of a former blue straggler in the cluster.

\subsection{EAGB18}

The power spectrum of EAGB18, the faintest EAGB star in our sample with a successful detection, exhibits a clear and well-defined power excess at $\nu_{max}=10.166\pm0.272\,\mu{\rm Hz}$. This corresponds to a stellar mass of $M=0.649\pm0.053\,M_{\odot}$.

\end{document}